\documentclass[12pt]{article}
\usepackage{amsmath}
\usepackage{graphicx}
\newtheorem{theorem}{Theorem}

\usepackage{enumerate}
\usepackage{url}

\usepackage{natbib}

\begin{document}


\title{A Theory of Individualism, Collectivism  and  Economic Outcomes}
\author{Kartik Ahuja, Mihaela van der Schaar and William R. Zame\thanks{Ahuja: Department of Electrical Engineering, UCLA, Los Angeles, CA 90095; ahujak@ucla.edu. van der Schaar:
Department of Electrical Engineering, UCLA, Los Angeles, CA 90095; mihaela@ee.ucla.edu. Zame: Department of Economics, UCLA, Los Angeles, CA 90095; zame@econ.ucla.edu.
We are grateful to John Asker, Moshe Buchinsky, Dora Costa and a seminar audience at UCLA for helpful comments.  Research support to Ahuja and van der Schaar was provided by the U.S. Office of Naval Research Mathematical Data Science Program; additional support to Ahuja was provided by the Guru Krupa Foundation. Any opinions, findings, and conclusions or recommendations expressed in this material are those of the authors and do not necessarily reflect the views of any funding agency.}}
\maketitle

\thispagestyle{empty}

\pagebreak

\begin{abstract}
This paper presents a dynamic model to study the impact on the economic outcomes in different societies during the Malthusian Era of individualism (time spent working alone)  and collectivism (complementary time spent working with others). The model is driven by opposing forces: a greater degree of collectivism provides a higher safety net for low quality workers but a greater degree of individualism allows high quality workers to leave larger bequests. The model suggests that more individualistic societies display smaller populations, greater per capita income and greater  income inequality.  Some (limited) historical evidence is consistent with these predictions.
\end{abstract}

\maketitle

 It is widely agreed (see \cite{landes1998wealth} for instance) that culture has an important influence on social outcomes and economic outcomes but there is little agreement on which aspects of culture are important for which economic outcomes,  whether these aspects are different in different eras, and  through what mechanisms culture operates.  This paper focuses on the impact of one aspect of culture -- the degree of  individualism vs. collectivism -- on the population, income and income distribution of societies in the period between the Neolithic Revolution and the Industrial Revolution -- a period in which life was (in the words of Thomas Hobbes) ``nasty, brutish and short,''  in which agriculture was the mainstay of economic activity and societies were stuck in the Malthusian trap, which \cite{Clark1}, \cite{Clark2}, \cite{Galor1} and \cite{Galor2} (and others) characterize by subsistence with no technological progress and little or no growth in either population or income.  We provide and analyze a model of the mechanism through which individualism and collectivism act.  Our model predicts that societies that are more individualistic (less collectivistic) tend to have smaller populations,  higher mean incomes, and greater income inequality.  Perhaps surprisingly, our model predicts that technological differences may matter a great deal for the size of the population but not for income or income inequality.  We offer some historical evidence that is consistent with the predictions of the model.(\cite{Clark1}, \cite{Clark2},  \cite{Galor1} and \cite{Galor2} have offered mathematical models of the Malthusian trap, but these  models do not offer an explanation of how or why cross-cultural differences -- in particular, differences in the degree of individualism and collectivism -- might have influenced outcomes in this period.  This is precisely the explanation our mathematical model is designed to provide. \cite{Roland1}, \cite{Roland2}, \cite{Roland3} offer an analysis of the impact of indivualism vs. collectivism in the era {\em after} the Industrial Revolution.  They argue that individualism rewards status and hence promotes innovation which in turn promotes growth.  However it does not seem that this explanation can explain the impact of individualism vs. collectivism in the Malthusian Era -- in which there was no growth.)

 We follow \cite{Hofstede} in viewing individualism as an aspect of culture that is associated with traits like acting independently and taking care of oneself and collectivism as an aspect  that is associated with mutual dependence amongst the members of the group. As in  \cite{Hofstede}, we view individualism and collectivism as aspects of the culture of a society,  which might or might not arise as aspects of the political structure.  We formalize the degree of individualism as the fraction of time that members of society work by themselves and enjoy only the output of their own activity and the degree of collectivism as the complementary fraction  of time that members of society work together and enjoy the output of the group activity.  We take these fractions as a universal social norm that is observed  by all the members of society and not as choices of different members of society (but these fractions differ across societies).       The societal division of time/labor matters because individuals differ in ability (physical strength, skill, etc.).  When working individually, output  per unit time depends on the individual's ability; when working collectively, output per unit time depends on the average ability of society.  When working collectively the less able members of society produce more per unit time than when working alone -- so a greater degree of collectivism provides  a social ``safety net'' for the low ability members of society.  On the other hand, when working collectively the more able members of society produce less  per unit time than when working alone -- so  a greater degree of collectivism decreases the wealth of the high ability members of society and hence the bequests they leave (to new-borns) when they die.  Because  income from production and inheritance from bequests {\em both} affect the path of individual wealth and hence  lifespan, the degree of collectivism and the complementary degree of individualism create {\em opposing forces\/}; the balance of these forces (and others) plays out in a complicated and subtle way. 
 \section{Model}
 The features of the model that we develop here are intended to represent (some aspects of) steady-state outcomes of  societies in the Malthusian Era, in which (changing) technology does not play an important role.  
 
 Before giving a formal mathematical description of the model, we begin with an informal verbal description that expands on what we have already said in the Introduction.  We consider a world populated by a continuum of individuals of two types either Low quality or High quality.\footnote{Allowing for more quality levels would complicate the analysis without altering the qualitative conclusions.}  Time is continuous and the horizon is infinite.  The lifecycle of an individual is: 
 \begin{itemize}
 	\item individuals are born and come into an inheritance;
 	\item during their lifetimes, individuals consume and produce;
 	\item individuals die and leave a bequest for succeeding individuals.
 \end{itemize}
 While they are alive and producing, each individual spends a fraction of its time working alone and consuming the output of its individual production, and the complementary fraction of its time working with others and sharing (equally) in the joint production.  We interpret these fractions as  (proxies for) the {\em degree of individualism} and the {\em degree of collectivism} of the society.\footnote{For example,  \cite{leibbrandt2013rise} show that lake based fishing areas are more individualistic and involve more isolated work by the individuals, while sea based fishing areas are more collectivistic and involve more collective work by the individuals.}    We view these fractions as social norms which are the same across all individuals in the society, rather than as individual choices.  (We are agnostic about the origins of these social norms; one possibility is that they are imposed by a governmental structure but there are many other possibilities.)  When individuals work alone, their output depends on their own quality; when individuals work with others, their output depends on the average quality of society.  In both modes,  output is subject to congestion: productivity is less when the total population is greater.  (This congestion is an essential part of Clark's argument for why societies remain in the Malthusian trap and plays an important role in our model as well.)  During their lifetimes, individuals consume at a constant rate.  (We discuss alternative assumptions below)  Some individuals produce less than they consume and eventually consume their entire inheritance; at that point their wealth is zero and they die in poverty.  Individuals who do not die in poverty eventually die of natural causes.  Individuals who die with positive wealth leave that wealth as a bequest to the new-born.

 We now turn to the formal mathematical description. We consider a continuous-time model with a continuum of individuals.  Some individuals are of High quality and some are of Low quality; it is convenient to index quality by $Q = 0,1$ (Low, High).\footnote{The individuals in our model are productive adults, so we view their quality as fixed and not changing over their lifetimes.}  The state of society at each moment of time is described by the {\em population distributions\/} ${\mathcal P}_0, {\mathcal P}_1$; ${\mathcal P}_Q(x,t)$ is the population of individuals of quality $Q$ who have  wealth less than or equal to $x$ at time $t$.  The  {\em population of individuals of quality $Q$ at time $t$} is 
 $$
 P_Q(t)  = \lim_{x\to \infty} {\mathcal P}_Q(x,t) 
 $$ 
 Thus  the {\em  total population at time $t$} is
 $$
 P(t) = P_0(t) + P_1(t)
 $$
 and the {\em average quality at time $t$} is
 $$
 \bar{Q}(t) = [0 \cdot P_0(t) + 1 \cdot P_1(t)]/P(t) = P_1(t)/P(t)
 $$
 
 Individuals are born at the  constant  rate $\lambda_f$ and die natural deaths at the constant rate $\lambda_d$.\footnote{ \cite{Clark1} argues that the fertility rate is an increasing function of the wealth of society and that the death rate is a decreasing function of the wealth of society.  Those features could be incorporated into our model without changing the qualitative conclusions, although at the expense of substantial mathematical complication.} Half of all newborns are of High quality and half are of Low quality.  (The assumption that the proportions of new-borns of High and Low quality are constant is made only for simplicity: none of the qualitative results would change if we assumed that quality is partly inheritable, so that the proportions of High and Low quality newborns  depend on the current population.  The assumption of equal proportions is made only to simplify the algebra.)  As we discuss below, some individuals also die in poverty. 
 
 While they are alive, individuals produce and consume.  We assume that each individual spends a fraction $z$ of its time working alone and the remaining fraction $1-z$ working with others.  As noted, we identify $z$ with the {\em degree of individualism} of the society and $1-z$ as the {\em degree of collectivism}.  When an individual works alone its production depends on its own quality and is consumed entirely by the individual; when it works with others its production depends on the average quality of society (at the given moment of time) and is shared; in both modes, productivity is subject to congestion and so diminishes with increasing population.  For simplicity, we assume productivity is linear in quality so productivity of an individual of quality $Q = 0,1$ at a given time $t$ when population is $P(t)$ and average quality is $\bar{Q}(t)$ is 
 $[Q - cP(t)]$ when working alone and $[ \gamma \bar{Q}(t) - cP(t)]$ when working with others, where $\gamma$ is a parameter that represents the efficiency of group production.  (Our assumptions about functional forms are made for tractability; our assumption that low quality individuals working alone produce nothing is simply a normalization.  As we will see below, the essential point is that, when working alone, low quality individuals produce less than they consume so that their wealth decreases.  (The role of the parameter $\gamma$ will be discussed in greater detail below.)   Hence the {\em overall productivity}  of an individual of quality $Q = 0,1$ is 
 \begin{equation}  
 \begin{split}
 F_Q(t) & = z[Q - cP(t)]+ (1-z) [\gamma \bar{Q}(t) - cP(t)]  \\
 & = zQ + (1-z)\gamma \bar{Q}(t) - cP(t)
 \end{split} \label{eqn:productivity}
 \end{equation}
 We emphasize that $Q$ is the innate and fixed quality of the (adult) individual and that $z, 1-z$ are characteristics of the society, and not individual choices. 
 
 We assume each individual consumes at the constant (subsistence) rate $k$; for algebraic simplicity (only) we take $k =1/2$. Hence the rate of production net of consumption for an individual with quality $Q$ is  $F_Q(t)-1/2$. Individuals who die at time $t$ leave a fraction $\eta < 1$ of their wealth as an inheritance for individuals born at the same time $t$; the remaining fraction $1-\eta$ of this wealth is lost in storage.     We write $y(t_0)$ as the (common) inheritance of individuals who are born at time $t_0$.  So an individual of quality $Q$ born at time $t_0$ begins life with {\em wealth} $X_q(t_0) = y(t_0)$; and its wealth  changes  during its lifetime at the rate:
 \begin{equation}
 dX_Q(t)/dt   =  F_Q(t)  - 1/2  \label{eqn:wealth} 
 \end{equation}
 We stress that an individual's wealth may shrink or grow; if it shrinks, it may eventually shrink to $0$ before the individual dies of natural causes in which case the individual dies in poverty.  Of course individuals who die in poverty do not leave an inheritance.  In our analysis, we  show that the system has a unique non-degenerate steady state.  In this steady state, $dX_0(t)/dt< 0$ and $dX_1(t)/dt > 0$ so the wealth of low quality individuals shrinks and the wealth of high quality individuals grows; it follows that some low quality individuals die in poverty but no high quality individuals die in poverty.

 We have defined the  state of society at time $t$ in terms of the population distributions ${\mathcal P}_0, {\mathcal P}_1$; however in analyzing the evolution of society it is more convenient to work with the  {\em population densities} $p_0, p_1$.  By definition,  
 $$
 {\mathcal P}_Q(x,t) = \int_0^x p_Q(\hat{x},t) \, d\hat{x}
 $$
 Working with densities is more convenient because their  evolution is determined by the following evolution equations, which are based on the principle of mass conservation:
 \begin{eqnarray}
 \frac{\partial p_0(x,t)}{\partial t} + \frac{\partial  p_0(x,t)}{\partial x} [F_0(t)-1/2] &=& -\lambda_d p_0(x,t)  \nonumber \\
 \frac{\partial p_1(x,t)}{\partial t} + \frac{\partial  p_1(x,t)}{\partial x} [F_1(t)-1/2] &=& -\lambda_d p_1(x,t) 
 \label{mass_cons}
 \end{eqnarray}
 The first term on the left hand sides of these PDE (\ref{mass_cons}) represents the rate of change of the population density at a given wealth level and the second term is the divergence of the flux; the right hand sides represents the rate at which individuals die due to natural causes.  Note that neither deaths in poverty nor births appear in the evolution equations.  This is because deaths in poverty only occur at  $x = 0$ and births only occur at $x = y(t)$ (inheritance at time $t$); deaths in poverty and births enter into the behavior of the system as  ``boundary conditions'' at $0, x = y(t)$ (see the Appendix).  Note that these evolution equations are {\em coupled} because productivity of agents of each quality depends on the {\em total population} rather than on the population of the given quality.  Note too that the ``boundary'' $x = y(t)$ is {\em moving} because inheritance $y(t)$  is a function of the population distributions and hence depends on time.
 
 We summarize the life-span of an individual in Figure 1.
 \begin{figure}
 	\centering
 	\includegraphics[width=5in]{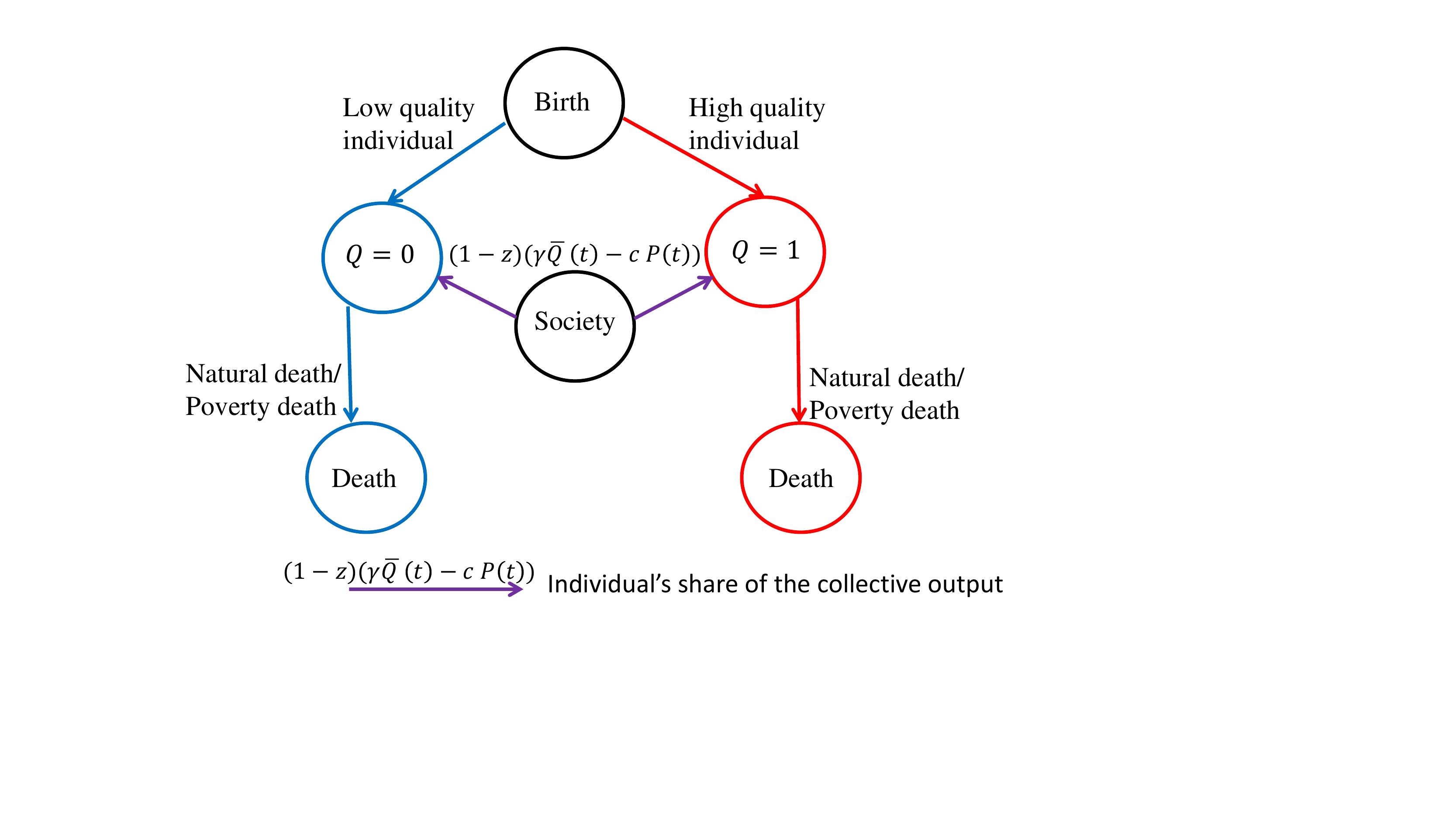}\caption{Individual's life span}
 \end{figure}


 \subsection{Steady State} We are interested in  societies in the steady state; because we are interested in the (long) period after the Neolithic Revolution and before the Industrial Revolution, during which there was little growth or change (see for instance  \cite{Clark1}), this seems reasonable.  We define the {\em steady state} as the state of the society in which the distribution of  individuals (of each type) across wealth levels is unchanging over time; i.e., $\partial p_Q(x,t)/\partial t \equiv 0$ for $Q = 0,1$.  In the the steady state,  the birth and (overall) death rate are constant and equal, so the populations $P_0(t), P_1(t), P(t)$ are constant; write $P_0^s, P_1^s, P^s$  for the steady state values.  Because the population is constant, so are the average quality $Q^s = P_1^s/P^s$,  the productivities of individuals of each quality $F_Q^s = zQ + (1-z)Q^s - cP^s$, and inherited wealth  $Y^s$. (All these values will be determined  endogenously by the  parameters of the model and the condition that the society is in steady state.)  
 
 There is always a degenerate steady state in which population is identically $0$.  In order to guarantee that a non-degenerate  steady state exists, we need four assumptions, which will be maintained in what follows without further comment.
 
 \bigskip
 
 \noindent {\bf Assumptions }
 \begin{enumerate}
 	\item $\lambda_d < \lambda_f<2\lambda_d$
 	\item $  \lambda_d/\lambda_f <\gamma$ 
 	\item $0<z < \frac{\lambda_d}{\lambda_f}$
 	\item $  1/(1+\frac{[1-\lambda_d/\lambda_f]}{\ln(\frac{1}{2[1-\lambda_d/\lambda_f]})})<\eta$
 \end{enumerate} 
 Some comments on these assumptions are in order.  If the natural birth  rate were less then the natural death rate then the population of society would shrink to $0$ in the long run so the only steady state would be degenerate.  Similar reasoning explains the second assumption.  To see why the third assumption is needed, suppose for a moment that $z = 0$, so that the society were completely collectivist.   In  a completely collectivist society, individual output depends only on average quality and not on individual quality, and hence net output in a steady state would be $\bar{Q}_s - cP_s - 1/2$.  If net output were positive, inheritance would blow up; if net output were negative, inheritance would shrink to $0$.  Hence in the steady state, net output must be $0$.  But this means that no individuals die in poverty; since the average quality of newly born agents is $1/2$,  the steady state average quality of the population must also be $1/2$ and  the steady state population must be $0$.  Hence a completely collectivist society cannot persist in a non-degenerate steady state.  Similar reasoning shows that an extremely individualistic society cannot persist in a non-degenerate steady state; the necessity of the given upper bound is derived in the proof of Theorem 1.  (Put differently: our model cannot apply to a society that is too collectivist or too individualistic.)  The last assumption asserts that the loss of wealth in inheritance is not too great.  (Recall that we have already assumed $\eta < 1$; i.e. {\em some} wealth is lost in inheritance.)  If $\eta$ were below the given bound then, as the proof of Theorem 1 demonstrates, the population of low quality individuals would go to $0$, which would once again be inconsistent with a non-degenerate steady state. 
  \begin{figure}
  	\centering
  	\includegraphics[trim={0 5cm 0 5cm 0}, clip, width=5in]{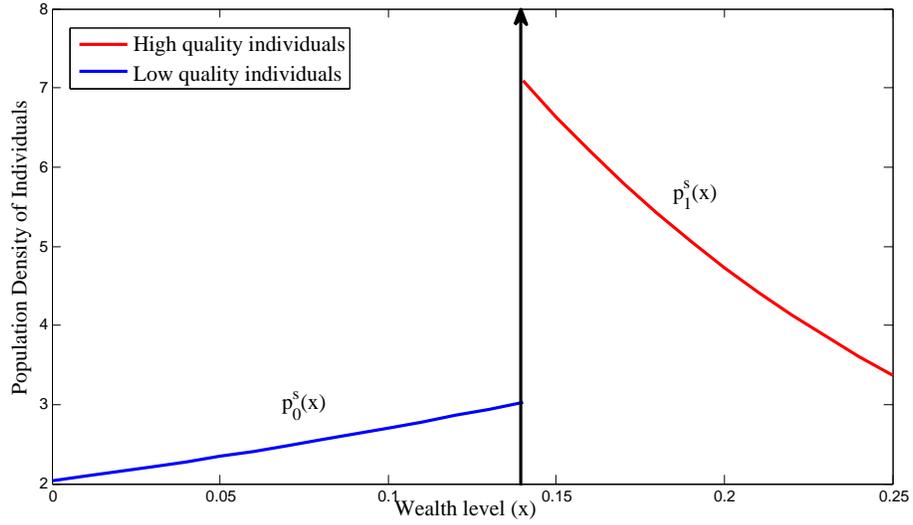}
  	\caption{Steady State Distribution of Wealth}
  \end{figure}

 Given these Assumptions, we can show that there is a unique non-degenerate steady state.  
 \bigskip
 
 \begin{theorem}
 	There is a unique non-degenerate steady state.
 \end{theorem}

 Figure 2 illustrates the steady state wealth distribution for typical values of the  parameters.  Note that the population of High quality individuals is greater than that of Low quality individuals, as indeed it must be given our assumptions.

 We defer the proof of this result (and all others) to the Appendix.
 
 \section{Model Predictions} 
 
 We now show that our model has strong -- and perhaps surprising -- implications for economic outcomes.  
 
 To understand what drives these implications, it is useful to think about the various forces at  work and how they manifest in the various aspects of the steady state.  Throughout the discussion, we take birth and death rates and inheritability $\eta$ of bequests as fixed, so that the steady state depends on the congestion coefficient $c$, the group efficiency $\gamma$ and the degree of individualism $z$. 
 
 The forces that these parameters generate can be seen most easily by comparing the non-degenerate steady state populations in different societies which differ in only one of these parameters.  With an obvious abuse of language we may speak of one of these parameters being or becoming larger.  Intuitively at least we can reason as follows.   
 \begin{itemize}
 	
 	\item If we hold group efficiency $\gamma$ and degree of individualism $z$ fixed then a larger congestion parameter $c$ generates a downward force on the population.  To see this, note that a larger $c$ implies a more negative congestion effect, so that productivity will be lower in both individual and group modes.  Hence  the wealth of low quality individuals will decline more quickly and wealth of high quality individuals will increase more slowly.  From this it also follows that individuals who die of natural causes will leave a smaller bequest, and hence that new-born individuals will come into a smaller inheritance.  In particular, low quality individuals will begin with less wealth, spend that wealth faster, and hence be more likely to die in poverty before they die of natural causes.  So if the congestion parameter is larger then the steady state population should be smaller.
 	
 	\item If we hold congestion $c$ and degree of individualism $z$ fixed then greater group efficiency $\gamma$ generates an upward force on the population.  To see this note that greater group efficiency means greater productivity for both high and low quality individuals when working with others.  Hence the wealth of low quality individuals will decline more slowly and the wealth of high quality individuals will increase more quickly.  From this, it also follows that individuals who die of natural causes will leave a larger bequest, and hence that new-born individuals will come into a larger inheritance.  In particular, low quality individuals will begin with greater wealth, spend that wealth more slowly, and hence be less likely to die in poverty before they die of natural causes.  So if group efficiency is greater then the steady state population should be larger.
 	
 	\item However if we hold congestion $c$ and group efficiency $\gamma$ fixed then a greater degree of individualism $z$ generates {\em both} upward {\em and} downward forces on the population.  To see this note that, on the one hand, low quality individuals  produce {\em more} per unit time when working with others than when working alone, so working with others provides low quality workers with a ``safety net.''  A greater degree of individualism lowers this  ``safety net'',  so that the wealth of low quality workers more quickly and they die in poverty {\em more often}.  On the other hand (at least if $\gamma$ is not too large)   high quality individuals produce {\em less} per until time when working with others than when working alone.  A greater degree of individualism therefore increases the rate at which high quality workers accumulate wealth, and hence increases the bequests they leave when they die, which in turn implies that  low quality individuals  begin life with greater wealth and tend to die in poverty {\em less often}.  Evidently, these forces work in opposite directions so the impact of the degree of individualism on population depends on the balance between them; we show below, the net effect depends on the relative magnitude of all the  parameters.  
 \end{itemize}
 As Theorem \ref{thm:population} below demonstrates formally, these intuitions about the impact of parameters on steady state population are indeed correct (and we can say even more about the impact of individualism).  However, we warn the reader that, as we will see later,  similar intuitions about the impact of parameters on other economic outcomes {\em are not correct\/}.  Although it may seem quite surprising, neither the congestion coefficient $c$ nor the group efficiency $\gamma$ influences mean income or income inequality.   
 \bigskip
 
 \begin{theorem}\label{thm:population}  In the non-degenerate steady state, population depends on $c, \gamma, z$ in the following way:
 	\begin{enumerate}[(a)]
 		\item $P^s$ is decreasing as a function of the congestion parameter $c$;
 		\item  $P^s$ is increasing as a function of group efficiency $\gamma$
 		\item for each $c > 0$ there is a threshold $\gamma^*$  such that
 		\begin{enumerate}[(i)]
 			\item if $\gamma < \gamma^*$ then $P^s$ is linearly increasing in $z$;
 			\item if $\gamma > \gamma^*$ then $P^s$ is linearly decreasing in $z$.
 		\end{enumerate}
 	\end{enumerate}
 \end{theorem}
 
 Theorem \ref{thm:population} describes the dependence of the total population on the various parameters but is silent about the dependence of the populations of each quality and the ratio of these populations.  Perhaps surprisingly, as Theorem \ref{thm:ratios} below asserts formally this ratio is  {\em independent} of all the parameters.  To understand the intuition for this conclusion, suppose the parameters change in such a way that the population of low quality workers grows.  Because the birth rate and the ratio of low quality births to high quality births are constant, the population of high quality workers must also grow -- and, as we show, it must grow at precisely the same rate as the population of low quality workers, so that the ratio of the populations remains constant.
 
 \bigskip
 
 \begin{theorem}\label{thm:ratios}  In the non-degenerate steady state, the population ratio $P^s_0/P^s_1$ is independent of $c, \gamma, z$.  
 \end{theorem}
 
 We now turn from population to income, in particular to mean income and to income inequality.  We identify income with output so the {\em mean income of society} in the steady state is 
 $$
 F^s = [F^s_0 P_0^s + F^s_1 P_1^s]/P^s
 $$
 
 \bigskip

 \begin{theorem}\label{thm:meanincome}
 	In the non-degenerate steady state, mean income is independent of $c, \gamma$ and linearly increasing in  the degree of individualism $z$.
 \end{theorem}
 
 At first glance, Theorem \ref{thm:meanincome} might seem startling.  It is natural to think of improved technology as manifested in a smaller congestion coefficient $c$ and a larger group efficiency $\gamma$; in view of Theorem \ref{thm:population} this would lead to an increase in the size of the population.  However as  population increases, so does congestion which reduces the (per capita) gains to the improved technology; in the steady state, these forces exactly balance out.  It seems important to point out that this is not simply an artifact of our model; Ashraf and Galor [3] argue that this is precisely what is observed in the data.   
 
 We measure income inequality in the familiar way as the Gini coefficient of the income distribution.  Because there are only two types of individuals, the Gini coefficient takes the particularly simple form
 $$
 F^s_1 P_1^s/F^s  P^s -P_1^s/P^s 
 = \left[\frac{P_1^s}{P^s} \right] \left[ \frac{F_1^s}{F^s} -1\right]
 $$

 \bigskip
 
 \begin{theorem}\label{thm:Gini} In  the non-degenerate  steady state the Gini coefficient is independent of $c, \gamma$ and increasing  in the level of individualism $z$.
 \end{theorem}

 \section{\hbox{Some Historical Evidence}}
 
 As we have said before, we intend our model to be descriptive of societies in the period between the Neolithic Revolution and the Industrial Revolution.   Although only a limited amount of  data is available for this period and there is some disagreement about its quality, it nevertheless seems appropriate to compare the predictions of our model with  the data that is available.  
 
 Our model makes use of a number of parameters: the birth and death rates $\lambda_f, \lambda_d$, the fraction $\eta$ of wealth that is inheritable, the coefficient $c$ of congestion, the group efficiency $\gamma$, and the degree $z$ of  individualism.  Unfortunately, none of these parameters can be observed directly.  (At least, none of these parameters were observed directly in the data that is available to us.)  What {\em is} available is an index of individualism calculated by \cite{Hofstede}, which we use as a proxy for $z$ (rescaled to lie in $[0,1])$.\footnote{A natural alternative would be to assume that $z$ is a (monotone) Box-Cox transformation \citep{Box-Cox} of Hofstede's index.  We have in fact computed the optimal Box-Cox transformation and carried through the regressions after performing the optimal Box-Cox  transformation; however,  there is almost no change in either the regression lines or the fit to the data.  The results of these regressions are available from the authors on request.}   In comparing the predictions of our model with  historical data we  make the simple (but perhaps heroic) assumption that birth and death rates and the fraction of wealth that is inheritable are the same across societies.  It seems completely implausible to assume that technologies are the same across societies -- and hence that the technological parameters $c, \gamma$ are the same across societies -- so we focus on the predictions for mean income and Gini coefficient, which are independent of these parameters.  
 
 To examine the implications of Theorem \ref{thm:meanincome} with historical data, we use estimates of GDP in 1500 CE  provided in \cite{Maddison} for Western Europe.   We identify mean income with GDP per capita     We use linear least-squares regression to compute the best-fitting straight line; the data and regression results can be seen in Figure 3.  (Note that some of the ``countries'' that appear in Figure 3 -- e.g. Italy -- did not exist in 1500.  Maddison uses the names to refer to the geographic areas  
 occupied by the {\em current} countries.)
 
 \begin{figure}
 	\includegraphics[trim={0cm 5cm 0cm 5cm},clip, width=5in]{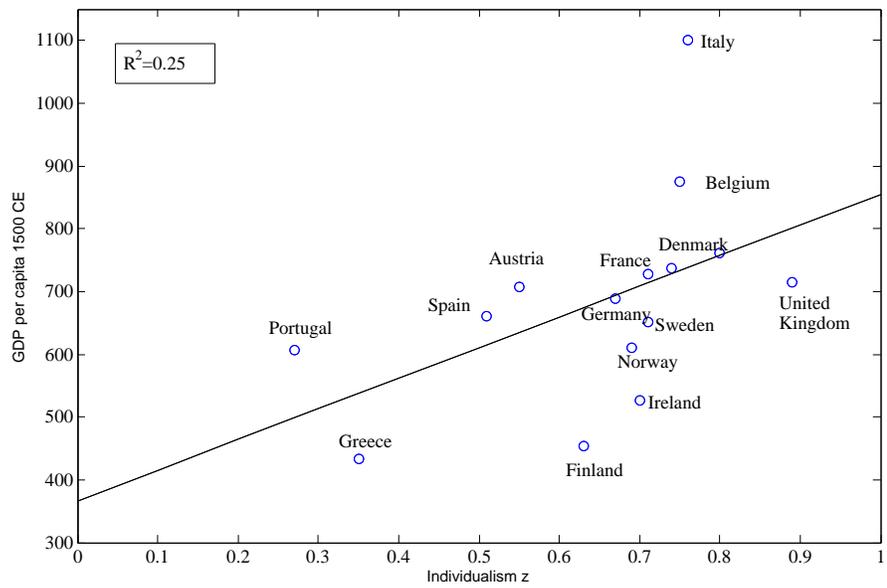}
 	\caption{Mean Income (GDP per capita) vs. Individualism}\label{fig:GDP}
 \end{figure}
 
 Unfortunately, we do not find any data for Gini coefficients from 1500 CE (the period of the data used above).  We therefore use the estimates of Gini coefficients given by    \cite{Williamson}.  This data from the (roughly) 100 year period 1788-1886 C.E. which might be thought to be after the Industrial Revolution and hence not appropriate for our model.  However, for those countries in which the Industrial Revolution arrived early (especially England, France and The Netherlands), the data and the calculations/estimations are from the beginning of this period, which would seem to be  predominantly  {\em before} the Industrial Revolution, while for those countries (especially Brazil, China and Peru) for which the data and the calculations/estimations are from the end of this period, the Industrial Revolution did not in fact arrive until much later.  The data and regression results can be seen in Figure 4.

 \begin{figure}
 	\includegraphics[width=6in]{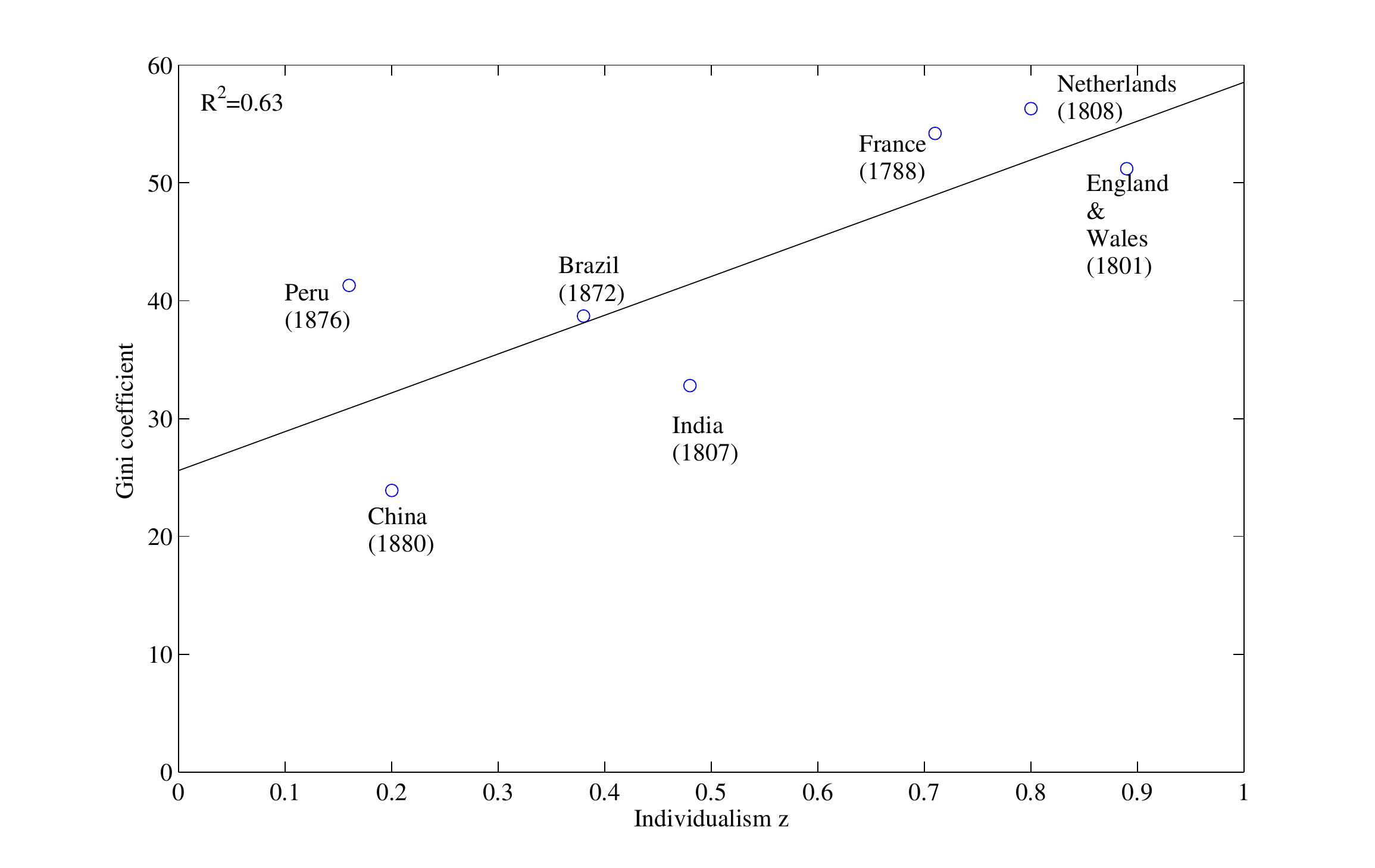}\caption{Gini Coefficients vs. Individualism}\label{fig:GINI}
 \end{figure}

 \section{Discussion and Conclusions }

 This paper proposes and analyzes a model that provides a mechanism by which the tension between individualism and collectivism can lead to different economic outcomes in different societies.  The  model captures  important features of the period between the Neolithic Revolution and the Industrial Revolution era as discussed in the work of  \cite{Clark1} and others. The model makes predictions about the impact of individualism and collectivism on different societies, and these predictions seem consistent with (limited) historical data.
 
 We reach no conclusion as to whether  individualism or collectivism is ``better'' -- indeed, the predictions of the model show that such a conclusion would depend entirely on the criteria used.  In particular, our prediction is that a greater degree of individualism leads to higher mean income (GDP per capita) but also to greater inequality; the first seems desirable, the second does not. 
 
 The  model presented above makes many simplifying assumptions -- but the model  could be generalized in many dimensions (allowing for non-linear congestion and non-constant fertility and death rates, for instance) without qualitative changes in the conclusions.  Other  generalizations might allow for the possibility that individual output and deaths due to poverty are stochastic (rather than deterministic) -- but such generalizations would seem to lead to enormous complications.
 
 We have confined our analysis to the steady state of the society which seems reasonable given that we are interested in the Malthusian Era in which there was little or no change.  However even in the Malthusian era there were shocks -- famines and epidemics -- which perturbed the system from its steady state, so it would certainly be of interest to know if the steady state of our model is at least locally stable -- i.e. if the system converges to the steady state from any initial point close to the steady state.  Unfortunately, this is an extremely complicated problem and well beyond or capabilities.  Out of the steady state the dynamics of our model are governed by a coupled pair of PDE's with a moving boundary constraint (and so the future evolution of the system depends on the entire wealth distribution and not just on a few aggregates).  Such dynamical systems are well-known to be extremely difficult to analyze -- or indeed, even to simulate numerically (because the numerical simulations can be extremely sensitive to the precise small details of the numerical approximation).

 Finally, the methodology proposed here suggests ways to think about contemporary societies as well -- although the analysis of contemporary societies will surely be more complicated because of rapidly changing technology, growing populations and trade.

\bibliographystyle{aea}
\bibliography{AER_final}

\appendix

\section{Mathematical Appendix}

Here we present the proofs for the formal results  discussed in the text.  Before we being, recall that the productivity of an individual of quality $Q$ at time $t$ is: 
$$
F_Q(t)= z[Q-cP(t)] + (1-z)[\bar{Q}(t)\gamma-cP(t)]=
zQ+(1-z)\gamma \bar{Q}(t) - cP(t)
$$
Note low quality individuals are always more productive when working collectively, but whether high quality individuals are more or less productive when working collectively depends on whether $\gamma\bar{Q}(t) > 1$ or $\gamma\bar{Q}(t) < 1$, and this is determined endogenously.

\noindent \textbf{Proof of Theorem 1 }  Since the proof is a bit roundabout, it may be useful to begin with an overview.  By definition, a steady state is a pair of density functions $p_0(x,t), p_1(x,t)$ that satisfy the evolution equations and are independent of time $t$.  In the steady state, the populations $P^s_0, P^s_1$  and inheritance $Y^s$ are constant, so average quality $Q^s$ and productivities $F^s_0, F^s_1$ are also constant.  Hence we can identify a steady state as a pair of functions $p_0(x), p_1(x)$ that satisfy the steady state evolution equations
\begin{equation}
\tag{SSEE0}\label{eqn:SSEE0}
\frac{\partial p_0(x)}{\partial x} [F^s_0 - 1/2] = -\lambda_d p_0(x) 
\end{equation}
\begin{equation}
\tag{SSEE1}\label{eqn:SSEE1}
\frac{\partial p_1(x)}{\partial x} [F^s_1 - 1/2] = -\lambda_d p_1(x) 
\end{equation}
and also satisfy the appropriate boundary conditions.  We therefore begin with candidate steady state populations $P^s_0, P^s_1$ and inheritance $Y^s$ (satisfying some conditions that must hold in any steady state of the system).  For any such triple, we show that the equations \ref{eqn:SSEE0}, \ref{eqn:SSEE1} admit unique solutions which yield the given steady state quantities.  We then show that the boundary conditions uniquely pin down the {\em unique}  triple of steady state quantities that correspond to an actual steady state of the society.

We begin by considering any non-degenerate solution $p_0^s(x), p_1^s(x)$ to the steady state evolution equations  (not necessarily satisfying boundary conditions).  From these, we  can derive the following steady state quantities:
\begin{itemize}
	
	\item the population of individuals with quality Q  
	\begin{equation}
	P_Q^{s}  =\int_0^{\infty} p_Q^s(x) dx   
	\end{equation}
	
	\item the total population 
	\begin{equation}
	P^s  = P_1^s+ P_0^s = \int_0^{\infty} [p_1^s(x) + p_0^s(x)] dx
	\end{equation}
	
	\item mean quality 
	\begin{equation}
	Q^s  = P_1^s/(P_0^s+P_1^s)
	\end{equation}
	
	\item productivity  of individuals of quality $Q$
	\begin{equation}
	F^s_Q = zQ+(1-z)\gamma Q^s-cP^s
	\end{equation}
	
	\item mean wealth 
	\begin{equation}
	X^s = \frac{\int_0^{\infty} x [p_1^s(x) + p_0^s(x) ] dx}{ \int_0^{\infty} [p_1^s(x) + p_0^s(x)]dx} 
	\end{equation}
	
	\item inheritance 
	\begin{equation}
	Y^s =  \lambda_d P^s X^s \eta / \lambda_f P^s = (\lambda_d/\lambda_f)X^s\eta
	\end{equation}
	
\end{itemize}
Because we have assumed a non-degenerate steady state we must have $P^s \not= 0$ so $P^s_0 \not= 0$ and $P^s_1 \not= 0$.  Note that  the three quantities  $P_0^s, P_1^s, Y^s$  determine all the others.   

We assert that in a non-degenerate steady state we must have $F_0^s < 1/2 < F_1^s $.  (Low quality individuals produce less than they consume; high quality individuals consume less than they consume.)  To show this we show that the other possibilities are incompatible with a non-degenerate steady state.  Note first of all that the definitions and the assumption that $0 < z < 1$ imply that $F_0^s < F_1^s$ so  we must rule out the only two other possibilities:
\begin{itemize}
	\item   $1/2 \leq F_0^s < F_1^s$.  If this were the case then the wealth of low quality individuals would be non-decreasing during their lifetimes and the wealth of high quality individuals would be strictly increasing during their lifetimes, so social wealth would be strictly increasing, which is impossible in the steady state.
	
	\item   $F_0^s < F_1^s \leq 1/2$.    If this were case then the wealth of low quality individuals would be strictly decreasing and the wealth of  high quality individuals would be non-increasing, so social wealth would be strictly decreasing, which is impossible in the steady state.
	
\end{itemize}
We therefore conclude that $F_0^s < 1/2 < F_1^s$ as asserted.

In order to show that a non-degenerate steady state of the society exists and is unique we proceed in the following way.  We have shown that, beginning with a solution $p_0^s, p_1^s$ to the steady state evolution  equations (\ref{eqn:SSEE0}), (\ref{eqn:SSEE1}) we can derive a triple of steady state quantities $P_0^s, P_1^s, Y^s$ having the property that $F_0^s < 1/2 < F_1^s $.  The first part of the proof is to show that, for every such triple of steady state quantities there is a unique  solution $p_0^s, p_1^s$ to the steady state evolution equations that yields the given steady state quantities.  The second part of the proof is to show that the boundary conditions uniquely pin down the triple of steady state quantities that correspond to an actual steady state of the society.

To this end, fix a triple of steady state quantities $P_0^s, P_1^s, Y^s$ for which   total population is positive $P_1^s+P_0^s=P^s>0$, inheritance is non-negative $Y^s\geq 0$ and for which the derived quantities $F_0^s, F_1^s$ satisfy $F_0^s < 1/2 < F_1^s $.  In any solution of the steady state evolution equations  that yields these steady state quantities, it is by true by definition that all individuals are born with the inheritance $Y^s$.  Because $F_0^s < 1/2 < F_1^s$, the wealth of low quality individuals is strictly decreasing while they are alive and the wealth of high quality individuals is strictly increasing while they are alive.  Hence,  $p_0^s(x)=0$ for $x > Y^s$ and $p_1^s(x) = 0$ for  $x < Y^s$; equivalently, $p_0^s$ is supported on $[0,Y^s]$ and $p_1^s$  is supported on $[Y^s, \infty)$.   From these facts we can determine the desired population distributions $p_0^s$ and $p_1^s$.

To determine $p_1^s$, set  $\lambda_1 = \lambda_d/[F_1^s-1/2]$.  For $x > Y^s$, the function $p_1^s$ solves the ODE: 
\begin{equation}
\frac{d p_1^s(x)}{d x} = -\lambda_1 p_1^s(x) 
\end{equation}
The solution to this  ODE is of the form	
\begin{equation}
p_1^s(x)=C_1 e^{-\lambda_1 (x-Y^s)}
\end{equation}
where the multiplicative constant $C_1$ is determined by initial conditions. Given $p_1^s$ we find that $P_1^s = C_1/\lambda_1$ so that  
\begin{equation}
p_1^s(x)= \begin{cases} P_1^s \lambda_1 e^{-\lambda_1 ( x-Y^s)} \ \  \mbox{ for } x> Y^s \\
0 \;\;\;\;\;\;\;\;\;\;\;\;\;\;\;\;\;\;\;\;\;\;  \ \, \mbox{ for } x<Y^s
\end{cases} \label{p1_exp}
\end{equation} 
Note that $\lambda_1= \lambda_d/[F_1^s-1/2]$ and recall that $F_1^s$ can be expressed in terms of $P_1^s,P_0^s$.

To determine $p_0^s$, set  $\lambda_0 = -\lambda_d/[F_0^s-1/2]$.  For $x <Y^s $ the function $p_0^s$ satisfies the  ODE:
\begin{equation}
\frac{d p_0^s(x)}{d x} = \lambda_0 p_0^s(x)    
\end{equation} 
The solution to this ODE is of the form	
\begin{equation}
p_0^s(x)=C_0 e^{\lambda_0 (x-Y^s)}
\end{equation}
where the multiplicative constant $C_0$ is  determined by  initial conditions. Given $p_0^s$ we find  that $P_0^s = (C_0/\lambda_0)(1-e^{-\lambda_0 Y^s})$ so that 
\begin{equation} 
p_0^s(x)=  \begin{cases} \left[P_0^s\lambda_0 /(1-e^{-\lambda_0 Y^s})\right] e^{\lambda_0 (x-Y^s)} \;\; \mbox{ for } x< Y^s\\
0 \;\;\;\;\;\;\;\;\;\;\;\;\;\;\;\;\;\;\;\;\;\;\;\;\;\;\;\;\;\;\;\;\;\;\;\;\;\;\;\;\;\;\;\;\; \mbox{ for } x>Y^s
\end{cases} \label{p0_exp}
\end{equation}
Note that $\lambda_0=-\lambda_d/[F_0^s-1/2]$ and recall that $F_0^s$ can be expressed in terms of 
$P_1^s,P_0^s$. 

By construction, the functions $p_0^s, p_1^s$ satisfy the steady state evolution equations.  Direct calculation shows that the steady state quantities derived from $p_0^s, p_1^s$  are precisely the quantities $P_0^s,P_1^s, Y^s$ with which we began.  This completes the first part of the proof.

We now turn to the second part of the proof which is to pin down the steady state values  of $P_1^s, P_0^s, Y^s$ that correspond to the (unique) non-degenerate steady state of the society.

Note first that because half of newborns are of low quality and half are of high quality, we have the following boundary condition:	
\begin{equation}
\lim_{x \downarrow Y^s} p_1^s(x)|F_1^s-1/2|
= \lim_{x \uparrow Y^s} p_0^s(x)|F_0^s-1/2|
\end{equation}
(As usual, $\lim_{x \downarrow Y^s}$ is the limit from the right and  
$\lim_{x \uparrow Y^s}$  is the limit from the left.)  
Simplifying yields
\begin{equation} 
P_1^s=\frac{P_0^s}{1-e^{-\lambda_0Y^s}}
\label{init_cond_steadystate}
\end{equation}
and hence  that 
\begin{equation}
e^{-\lambda_0Y^s} = (2-P^s/P_1^s) \label{exp_relP}
\end{equation}

Next we compute the rate  $\mu^s$ at which individuals die in poverty in the steady state.  (Of course, only low quality individuals die in poverty.) 
\begin{equation}
\begin{split}
\mu^s &= f_0^s(0)|F_0^s-1/2| \\
&= C_0e^{-\lambda_0 Y^s}|F_0^s-1/2| \\
&= (C_0/\lambda_0) \lambda_d e^{-\lambda_0 Y^s} \\
&= P_1^s \lambda_d e^{-\lambda_0Y^s} \\
& = (2P_1^s-P^s)\lambda_d 
\label{mu_s}
\end{split}
\end{equation}

In the steady state the population is constant so the birth rate must equal to death rate, which yields the second boundary condition:
\begin{equation} (\lambda_f P^s-\lambda_d P^s-\mu^s) =0\end{equation}
Substituting gives: 
\begin{equation} \lambda_fP^s- \lambda_d P^s -\lambda_d (2P_1^s-P^s)=0\end{equation}
Hence, we have 
\begin{equation}
P_1^s = \lambda_f/(2\lambda_d) P^s   \label{eqn:populationratios}
\end{equation}

By assumption,  $\eta <1$ is the fraction of wealth that is transferred across generations so: 
\begin{equation} \lambda_f Y^s= \eta \lambda_d X^s \label{wealth_tx}
\end{equation}

Next we compute  $X^s$.
\begin{equation}
X^s = \frac{1}{2-e^{-\lambda_0 Y^s}}\Big[\int_0^{Y^s} \lambda_0 x e^{\lambda_0(x-Y^s)} dx  +  \int_{Y^s}^{\infty} \lambda_1 x e^{-\lambda_1(x-Y^s)} dx\Big] 
\end{equation}
Integration by parts yields:
\begin{align}
\begin{split}
\int_0^{Y^s} \lambda_0 x e^{\lambda_0(x-Y^s)} dx &= \Big[\frac{e^{-\lambda_0 Y^s}-1}{\lambda_0}  \\ 
& \ \ \ \ \ \ + \;\;\;\frac{\lambda_0Y^s e^{-\lambda_0 Y^s}}{\lambda_0} +Y^s-Y^se^{-\lambda_0 Y^s} \Big] \\
\int_{Y^s}^{\infty} \lambda_1 x e^{-\lambda_1(x-Y^s)} dx &=
\frac{1+\lambda_1Y^s}{\lambda_1}
\end{split}
\end{align}
We use the above expressions to simplify $X^s$:
\begin{equation}
X^s= 2\frac{P_1^s}{P^s}Y^s + \frac{1}{\lambda_d}(\frac{P_1^s}{P_s}\left[z+(1-z)\gamma\right]) -\frac{1}{\lambda_d}(cP^s+1/2)
\end{equation}
We can substitute $\frac{P_1^s}{P^s}$ from \eqref{eqn:populationratios} in the above and substitute $X^s$ from \eqref{wealth_tx}  to obtain
\begin{equation}
Y^s =  \left(\frac{\lambda_f}{2\lambda_d}\left[z+(1-z)\gamma\right]-cP^s-1/2\right)\left(\frac{\eta}{\lambda_f (1-\eta)}\right) \label{Ys_f}
\end{equation}

Using the  equations  \eqref{Ys_f}, \eqref{eqn:populationratios} and \eqref{exp_relP} we will determine each of the desired quantities.
We write \eqref{exp_relP} as follows. 
\begin{equation}
e^{-\lambda_0 Y^s}  = 2-P^s/P_1^s 
\end{equation}
Substitute \eqref{eqn:populationratios} and the expression for $\lambda_0$ in the above and then take logarithms to obtain:
\begin{equation}
\lambda_dY^s   =  \ln \left[2-\frac{2\lambda_d}{\lambda_f} \right]  \left[(1-z)\gamma\frac{\lambda_f}{2\lambda_d}-cP^s-1/2)\right] 
\end{equation}
Substitute $cP^s+1/2$ from \eqref{Ys_f} in the above to obtain:
\begin{equation}
\lambda_d Y^s  = \ln \left[2-\frac{2\lambda_d}{\lambda_f}\right]  
\left((1-z)\gamma\frac{\lambda_f}{2\lambda_d}+\left[\frac{\lambda_f(1-\eta)}{\eta} \right]Y^s -\frac{\lambda_f}{2\lambda_d}(z+(1-z)\gamma)\right)
\end{equation}
We can simplify the above to obtain the  final expression for $Y^s$: 
\begin{equation}
Y^s = \frac{\eta z}{2\lambda_d(1-\eta +\beta)} \label{Ys_final}
\end{equation}
where $\beta=-[\eta \,\lambda_d/\lambda_f]/ \ln[2-\frac{2\lambda_d}{\lambda_f}]$.
In a non-degenerate steady state we must have $X^s>0$. We know $X^s =\frac{\lambda_f Y^s}{\lambda_d \eta}$; since $\lambda_d<\lambda_f<2\lambda_d $ it follows that $ 2(1-\frac{\lambda_d}{\lambda_f}) \in (0,1)$  and hence that $(1-\eta + \beta)>0$ and that $X^s>0$ as required.

Now we substitute \eqref{Ys_final} in \eqref{Ys_f} to obtain the expression for $P^s$ as follows.
\begin{equation}
cP^s =\frac{\lambda_f}{2\lambda_d}\left[\gamma+z\left(\frac{\beta}{1-\eta+\beta}-\gamma\right)\right]-\frac{1}{2}\label{P_s}
\end{equation}

Since $\lambda_f>\frac{\lambda_d}{\gamma}$ the above expression is greater than zero when $z=0$ and since $\eta >  1/(1+\frac{[1-\lambda_d/\lambda_f]}{\ln(\frac{1}{2[1-\lambda_d/\lambda_f]})})$ the above expression is greater than zero when $z=1$. This ensures that $P^s>0$. 
We know that 
\begin{equation}
P_1^s=\left[\frac{\lambda_f}{2c\lambda_d}\right] \left[\frac{\lambda_f}{2\lambda_d}\left[\gamma+z\left(\frac{\beta}{1-\eta+\beta}-\gamma\right)\right]-\frac{1}{2}\right]\label{P1_final}
\end{equation}
and
\begin{equation}
P_0^s=\left[\frac{1}{c}-\frac{\lambda_f}{2c\lambda_d}\right] \left[\frac{\lambda_f}{2\lambda_d}\left[\gamma+z\left(\frac{\beta}{1-\eta+\beta}-\gamma\right)\right]-\frac{1}{2}\right]\label{P0_final}
\end{equation}

Since $P^s>0$ both $P_1^s$ and $P_0^s$ are greater than zero.
This derivation was based on the assumption that $F_0^s<1/2 < F_1^s$; we now check that this is indeed true for the derived values of $P_0^s, P_1^s, X^s$.

We treat $F_1^s$ first. Substitute $\eqref{P_s}$ to obtain
\begin{align}
\begin{split}
F_1^s-1/2 &= z + (1-z)\gamma\frac{\lambda_f}{2\lambda_d} -cP^s -1/2  \\
& = z(1-\frac{\lambda_f}{2\lambda_d}) +  \frac{ z\lambda_f(1-\eta)}{2\lambda_d (1-\eta +\beta) }\label{F_1}\\
\end{split}
\end{align}
Because $z > 0$ and $\lambda_f<2\lambda_d$ the first term in the right hand side is strictly positive.   Because  $(1-\frac{\lambda_f}{2\lambda_d})>0$  and $(1-\eta+\beta)>0$ the second term is strictly positive, so  $ F_1^s-1/2>0$.


We now turn to $F_0^s$.  Substitute $\eqref{P_s}$ to obtain
\begin{align}
\begin{split}
F_0^s-1/2 & = (1-z)\gamma\frac{\lambda_f}{2\lambda_d} - cP^s -1/2 \\
& =z\left[\frac{\lambda_f}{2\lambda_d}\right] \left[ \frac{-\beta}{1-\eta+ \beta}\right] \label{F_0}
\end{split}
\end{align}
Since $z>0$,  $\beta>0$ and $(1-\eta+\beta)>0$, we conclude that $F_0^s-1/2 < 0$.  To see that 
$F^s_0 > 0$ we calculate:

Now we have determined the values of $P_1^s, P_0^s, Y^s$ in \eqref{P1_final}, \eqref{P0_final} and  \eqref{Ys_final}. We can substitute these in  \eqref{p1_exp} and \eqref{p0_exp}  to obtain the final  distribution function.  This completes the proof.   

\bigskip

\noindent \textbf{Proof of Theorem 2 } In the proof of Theorem 1 we arrived at an expression for $cP^s$ in \eqref{P_s}, so we conclude that  
\begin{equation}
P^s =\left\{\frac{1}{c}\right\} \left\{\frac{\lambda_f}{2\lambda_d}\left[\gamma+z\left(\frac{\beta}{1-\eta+\beta}-\gamma\right)\right]-\frac{1}{2}\right\}
\end{equation}
where $\beta=-[\eta \,\lambda_d/\lambda_f]/ \ln[2-\frac{2\lambda_d}{\lambda_f}]$.  It is immediate that $P^s$ is decreasing in $c$ and increasing in $\gamma$.  $P^s$ is evidently  linear in $z$; $P^s$ is decreasing in $z$ if $\gamma > \frac{\beta}{1-\eta+\beta}$ and is  increasing in $z$ if  $\gamma < \frac{\beta}{1-\eta+\beta}$, as asserted.  

\bigskip

\noindent \textbf{Proof of Theorem 3 } In the proof of Theorem 1, we arrived at  equation (\ref{eqn:populationratios}) which expresses the population $P^s_1$ of high quality individuals as a fraction of the total population $P^s$.  Since $P^s = P^s_0 +P^s_1$, simple algebra shows that the steady state population ratio is
\begin{equation}
\frac{P^s_0}{P^s_1} = 1 - \frac{\lambda_f}{2\lambda_d}
\end{equation}
Note that Assumption 1 guarantees that the right hand side is strictly positive and less than 1. This completes the proof.

\bigskip

\noindent \textbf{Proof of Theorem 4 } We first derive the expression for mean income. We know that $P_1^s/P^s = Q^s = \frac{\lambda_f}{2\lambda_d}$. In the simplification given below we use the expression derived in \eqref{F_1} and \eqref{F_0}.
\begin{equation}
\begin{split}
F^s &= Q^sF_1^s + \left[1-Q^s\right]F_0^s 
\end{split}
\end{equation}
\begin{equation}
F^s  =\frac{(1-\eta)z}{1-\eta+\beta}+\frac{1}{2} \label{mean_income}
\end{equation}
Note that mean income $F^s$ is independent of the technological parameters $c, \gamma$ and linear in the degree of individualism $z$.  Because $\eta < 1$, mean income is increasing in the degree of individualism $z$.   

\bigskip

\noindent \textbf{Proof of Theorem 5 } We have seen in the proof of Theorem 1 that both income levels are positive, so writing $A=\frac{(1-\eta)}{(1-\eta+\beta)}$ and performing the requisite algebra yields a convenient expression for the Gini coefficient is:

\begin{align}
\begin{split}
Gini & = \frac{Q^s F_1^s }{Q^s F_1^s + \left(1-Q^s\right)F_0^s} - Q^s \\
& = Q^s\left[\frac{(1-Q^s)F_1^s-(1-Q^s)F_0^s}{F^s}\right] \\
& =Q^s(1-Q^s)\left(\frac{F_1^s-F_0^s}{F^s}\right) \\
& = Q^s(1-Q^s) \left( \frac{z}{Az+1/2} \right)
\end{split}
\end{align}

Because the steady state average quality $Q^s$ depends only on the steady state population ratio $P^s_0/P^s_1$, which is independent of the technological parameters $c, \gamma$, we see that the Gini coefficient is also independent of the technological parameters $c, \gamma$.  

Finally, differentiating the expression for the Gini coefficient yields
\begin{align}
\begin{split}
\frac{\partial Gini}{\partial z} = \frac{\lambda_f}{2\lambda_d}\left[1-\frac{\lambda_f}{2\lambda_d}\right] \left[ \frac{1}{2(Az+1/2)^2}\right] 
\end{split}
\end{align}
Since $\lambda_f<2\lambda_d$ the Gini coefficient is increasing in the level of individualism $z$, as asserted.

\end{document}